# Current Conveyor Based Multifunction Filter

*Manish Kumar*
Electronics and Communication Engineering Department
Jaypee Institute of Information Technology
Noida, India

*M.C. Srivastava*
Electronics and Communication Engineering Department
Jaypee Institute of Information Technology
Noida, India

*Umesh Kumar*
Electrical Engineering Department
Indian Institute of Technology
Delhi, India

*Abstract*—The paper presents a current conveyor based multifunction filter. The proposed circuit can be realized as low pass, high pass, band pass and elliptical notch filter. The circuit employs two balanced output current conveyors, four resistors and two grounded capacitors, ideal for integration. It has only one output terminal and the number of input terminals may be used. Further, there is no requirement for component matching in the circuit. The parameter resonance frequency ($\omega_0$) and bandwidth ($\omega_0/Q$) enjoy orthogonal tuning. The complementary metal oxide semiconductor (CMOS) realization of the current conveyor is given for the simulation of the proposed circuit. A HSPICE simulation of circuit is also studied for the verification of theoretical results. The non-ideal analysis of CCII is also studied. *(Abstract)*

## I. INTRODUCTION (HEADING 1)

Active filters are widely used in the signal processing and instrumentation area. The well known advantage of current mode operation, such as better linearity, simple circuitry, low power consumption and greater bandwidth becomes more attractive as compared to voltage-mode counterpart with introduction of Current Conveyor II(CCII). The application and advantages in the realization of various active filters using current conveyors has received considerable attention [1]-[5]. Some voltage mode multifunction filter using current conveyors have also been proposed. In 1995 Soliman [1] proposed Kerwin-Huelsman-Newcomb (KHN) biquad with single input and three outputs, which realizes low-pass, band-pass and high-pass filter. The circuit employs five current conveyor (CCII), two capacitors and six resistors. In 1997 Higahimura et al. [2] proposed a universal voltage-mode filter that can realize low-pass, high-pass, band-pass, all-pass and notch filter using seven current conveyors, two capacitors and eight resistors. Ozoguz et. al. [3] realized high-pass, low-pass and band-pass filter using three positive current conveyor and five passive components. In 1999 Chang and Lee [4] proposed voltage mode low-pass, band-pass and high-pass filter with single input and three outputs employing only current conveyors, two grounded capacitors and three resistors. Toker et. al. [5] realized high output impedance transadmittance type continuous time multifunction filter (low-pass, high-pass and band-pass) employing three positive type current conveyor and five passive components. The circuit proposes high output impedance.

In this paper a circuit employing two balanced output current conveyors, four resistors and two grounded capacitors is proposed. This circuit has one output terminal and four input terminals. All the basic filters (low pass, high pass, band pass and notch filter) may be realized by selecting appropriate input terminals of the circuit.

The following section presents circuit description of the balanced output current conveyor. The sensitivity analysis, nonideal analysis of balanced output current conveyors, simulation results and conclusion are discussed in the subsequent sections.

## II. CIRCUIT DESCRIPTION

The balanced output current conveyor is shown in fig 1 with its symbol, characterized by the port relations as given by "(1)"

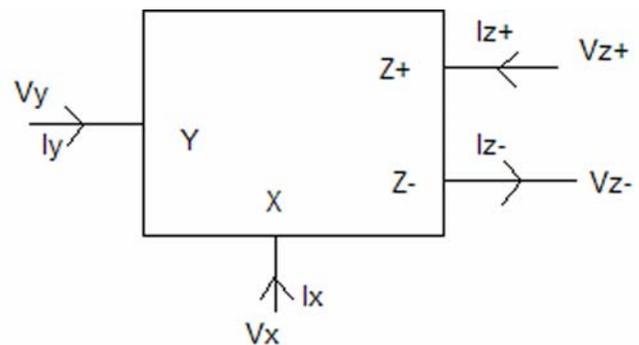

Figure 1. Symbol of balanced output current Conveyor

$$\begin{bmatrix} V_x \\ I_y \\ I_{z+} \\ I_{z-} \end{bmatrix} = \begin{bmatrix} 0 & B & 0 & 0 \\ 0 & 0 & 0 & 0 \\ K & 0 & 0 & 0 \\ -K & 0 & 0 & 0 \end{bmatrix} \begin{bmatrix} I_x \\ V_y \\ V_{z+} \\ V_{z-} \end{bmatrix} \quad (1)$$

$$\omega_0 = \sqrt{\frac{R_1 + R_6}{R_1 R_4 R_6 C_2 C_5}} \quad (4)$$

$$\frac{\omega_0}{Q} = \frac{1}{R_2 C_3} \quad (5)$$

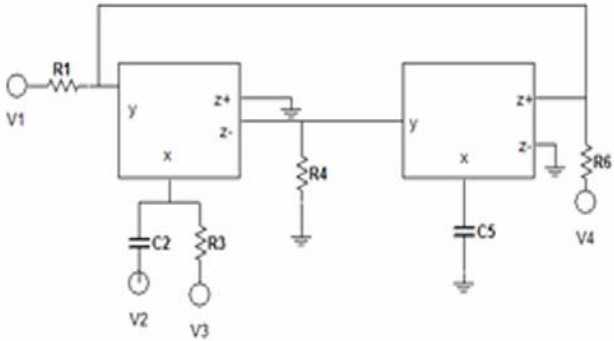

Figure 2. Proposed Voltage Mode Multifunction Filter

TABLE 1  VARIOUS FILTER RESPONSES

| Filter\Input | $V_1$ | $V_2$ | $V_3$ | $V_4$ |
|---|---|---|---|---|
| Low-pass | 1 | 0 | 0 | 1 |
| High-pass | 0 | 1 | 0 | 0 |
| Band-pass | 0 | 0 | 1 | 0 |
| Notch | 1 | 1 | 0 | 1 |

$$Q = R_3 \sqrt{\frac{C_2(R_1 + R_6)}{R_1 R_4 R_6 C_5}} \quad (6)$$

It can be seen from a perusal of "(4)" - "(6)" that both the center frequency and quality factor are independent. An inspection of "(4) and "(5)" shows that $\omega_0$ and $\omega_0/Q$ can be orthogonally tuned through $R_6$ and /or $C_5$ and $R_3$ and /or $C_2$ in that order.

The values of B and K are frequency dependent and ideally B=1 and K=1.

The proposed circuit shown in fig 2 employs only two balanced output current conveyor, four resistor and two capacitors. The grounded capacitors are particularly very attractive for the integrated circuit implementation.

## III. SENSITIVITY ANALYSIS

The sensitivity analysis of the proposed circuit is presented in terms of the sensitivity of $\omega 0$ and Q with respect to the variation in the passive components as follows:

$$V_{out} = \frac{s^2 C_2 C_5}{D(s)} \left[ \frac{R_3 R_6}{s^2 C_2 C_5} V_1 + R_1 R_3 R_4 R_6 V_2 + \frac{R_1 R_4 R_6}{s C_2} V_3 + \frac{R_1 R_3}{s^2 C_2 C_5} V_4 \right] \quad (2)$$

$$S^{\omega_0}_{C_2, C_5, R_4} = -\frac{1}{2} \quad (7)$$

$$S^{\omega_0}_{R_1} = -\frac{R_{61}}{2(R_1 + R_6)} \quad (8)$$

Where

$$D(s) = s^2 C_2 C_5 R_1 R_3 R_4 R_6 + s C_5 R_1 R_4 R_6 + (R_1 R_3 + R_3 R_6) \quad (3)$$

$$S^{\omega_0}_{R_6} = -\frac{R_1}{2(R_1 + R_6)} \quad (9)$$

Thus by using "(2)" we can realize low-pass, band-pass, high-pass and notch filter responses at the single output terminal by applying proper inputs at different node as shown in table1.

$$S^Q_{R_3} = 1 \quad (10)$$

The denominators for the all filter responses are same. The filtering parameter $\omega_o$, $\omega_o/Q$ and Q are given by

$$S^Q_{C_2} = \frac{1}{2} \quad (11)$$

$$S^Q_{R_4, C_5} = -\frac{1}{2} \quad (12)$$

$$S_{R_1}^Q = -\frac{R_{61}}{2(R_1 + R_6)} \quad (13)$$

$$S_{R_6}^Q = -\frac{R_1}{2(R_1 + R_6)} \quad (14)$$

As per these expressions, both the $\omega_0$ and Q sensitivities are less than ± ½ with a maximum value of $S_{R_3}^Q = 1$.

## IV. NONIDEAL ANALYSIS

Practically B and K are frequency dependent with dominant poles and therefore intended to nonideal. The non ideal behavior of the output $V_{out}$ may be expressed by "(15)".

$$V_{out} = \frac{s^2 C_2 C_5}{D(s)} \left[ \frac{R_3 R_6}{s^2 C_2 C_5} V_1 + K_1 K_2 B_2 R_1 R_3 R_4 R_6 V_2 + \frac{K_1 K_2 B_2 R_1 R_4 R_6}{s C_2} V_3 + \frac{R_1 R_3}{s^2 C_2 C_5} V_4 \right] \quad (15)$$

Where

$$D(s) = K_1 K_2 B_1 B_2 (s^2 C_2 C_5 R_1 R_3 R_4 R_6 + s C_5 R_1 R_4 R_6) + (R_1 R_3 + R_3 R_6) \quad (16)$$

$$\omega_0 = \sqrt{\frac{R_1 + R_6}{K_1 K_2 B_1 B_2 R_1 R_4 R_6 C_2 C_5}} \quad (17)$$

$$\frac{\omega_0}{Q} = \frac{1}{C_2 R_3} \quad (18)$$

$$Q = R_3 \sqrt{\frac{C_2(R_1 + R_6)}{K_1 K_2 B_1 B_2 R_1 R_4 R_6 C_2 C_5}} \quad (19)$$

It can be observed that the effect of non ideality behavior on cutoff frequency ($\omega_0$) and Q will be negligible and bandwidth will not be affected. The sensitivity analysis of cutoff frequency and the quality factor with respect to $K_1$, $K_2$, $B_1$ and $B_2$ are as follows:

$$S_{B_1,B_2,K_1,K_2}^{\omega_0} = -\frac{1}{2} \quad (20)$$

$$S_{B_1,B_2,K_1,K_2}^Q = -\frac{1}{2} \quad (21)$$

The $\omega_0$ and Q sensitivities with respect to passive components are same in case of ideal balanced output current conveyor.

TABLE 2    TRANSISTOR ASPECT RATIOS FOR BALANCED OUTPUT CCII

| Transistor | W(µm)/L(µm) |
|---|---|
| M1, M2 | 20/1 |
| M3 | 50/2.5 |
| M4, M6, M8 | 80/2.5 |
| M5 | 100/2.5 |
| M7,M9 | 50/2.5 |

## V. SIMULATION RESULT

The HSPICE simulation with 0.5µm CMOS transistor model provided by MOSIS has been carried out for the realization of balanced output current conveyor as shown in fig (3). Table 2 list the dimension on NMOS and PMOS transistor of this circuit. Figure 4 displays the simulation result for the proposed filter. The circuit is designed for $\omega_0$ = 14.14 KHz and Q=2 by considering $R_1 = R_4 = R_6 = 10$KΩ, $C_2 = C_5 = 10$nF and $R_3 = 14$KΩ. The theoretical results have been verified to match with simulation result.

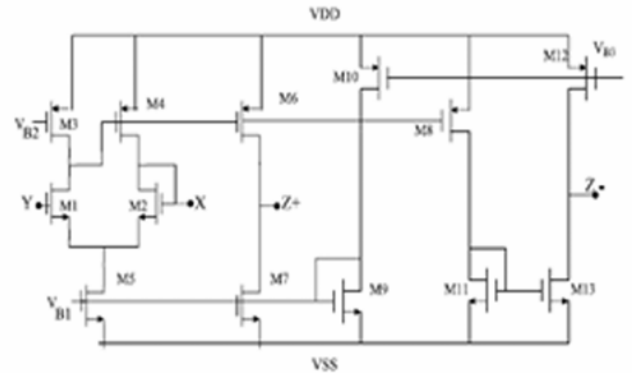

Figure 3.  CMOS Circuit for Balanced Current Conveyor II

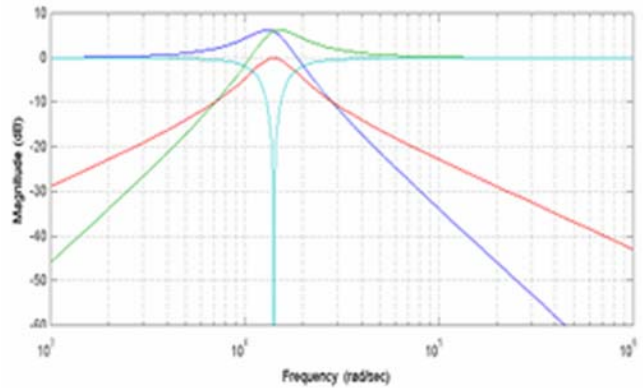

Figure 4.  Multifunction Filter Responce

*(IJCSIS) International Journal of Computer Science and Information Security,*
*Vol. 7, No. 2, 2010*

## VI. Conclusion

The circuit proposed by Hong *et al*. and Chang *et. al.* uses more active and passive components. Whereas the circuit proposed in this paper generates low-pass, high-pass, band-pass and notch filter using two balanced output current conveyors, four resistors and two capacitors. The circuit provides more number of filter realizations at the single output terminal using two current conveyors. In addition of this proposed circuit does not have any matching constraint/cancellation condition. The circuit employs' grounded capacitor, suitable for IC fabrication. The circuit enjoys the othogonality between the cutoff frequency and the bandwidth. It has low sensitivities figure of both active and passive components.
## References

[1] A. M. Soliman, "Kerwin–Huelsman–Newcomb circuit using current conveyors," Electron. Lett., vol. 30, no. 24, pp. 2019–2020, Nov. 1994.

[2] M. Higasimura and Y. Fukui, "Universal filter using plus-type CCII's," Electron. Lett. vol. 32, no. 9, pp. 810-811, Apr. 1996.

[3] S. Ozoguz, A. Toker and O. Cicekoglu, "High output impedance current-mode multifunction filter with minimum number of active and reduced number of passive elements," Electronics Letters, vol. 34, no 19, pp. 1807-1809, 1998

[4] Chun-Ming Chang and Ming- Jye Lee, "Voltage mode multifunction filter with single input and three outputs using two compound current conveyors," IEEE Trans. On Circuits and Systems-I: vol. 46, no. 11, pp.1364-1365, 1999.

[5] A. Toker, O. Çiçekoglu, S. Özcan and H. Kuntman ," High-output-impedance transadmittance type continuous-time multifunction filter with minimum active elements," International Journal of Electronics, Volume 88, Number 10, pp. 1085-1091, 1 October 2001.

[6] A. M. Soliman, "Current mode universal filters using current conveyors: classfication and review," Circuits Syst Signal Process, vol. 27, pp. 405-427, 2008.

[7] P. V. Anada Mohan, Current Mode VLSI Analog Filters, Birkhauser, Boston, 2003.

## AUTHORS PROFILE

Authors Profile …

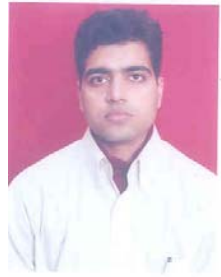

**Manish Kumar** was born in India in 1977. He received his B.E. in electronics engineering from S.R.T.M.U. Nanded in 1999 and M.E. degree from Indian Institute of Science, Bangalore in 2003. He is perusing is Ph.D. He is working as faculty in Electronics and Communication Engineering Department of Jaypee Institute of Information Technology, Noida He is the author of 10 papers published in scientific journals and conference proceedings. His current area of research interests includes analogue circuits, active filters and fuzzy logic.

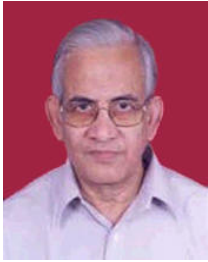

**M. C. Srivastava**  received his B.E. degree from Roorkee University (now IIT Roorkee), M.Tech. from Indian Institute of Technology, Mumbai and Ph.D from Indian Institute of Technology, Delhi in 1974.  He was associated with I.T. BHU, Birla Institute of Technology and Science Pilani, Birla Institute of Technology Ranchi, and ECE Dept. JIIT Sector-62 Noida. He has published about 60 research paper.  His area of research is signal processing and communications. He was awarded with Maghnad Saha Award for his research paper.

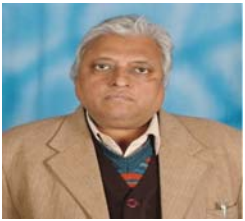

**Umesh Kumar** is a senior member, IEEE. He received B.Tech and Ph.D degree from IIT Delhi. He has published about 100 research papers in various journals and conferences. He is working as faculty in Electrical Engineering Department, IIT Delhi.